\documentclass[journal=jacsat,manuscript=article]{achemso}

\usepackage{chemformula} 
\usepackage[T1]{fontenc} 

\author{Siyuan Gao}
\affiliation[Keio University]
{Keio University, 3-14-1 Hiyoshi, Kohoku, Yokohama, Kanagawa, Japan.}
\email{gaosy@keio.ac.jp}
\author{Tianji Liu}
\affiliation[GPL Photonics Laboratory]
{GPL Photonics Laboratory, State Key Laboratory of Luminescence Science and Technology,
Changchun Institute of Optics, Fine Mechanics and Physics, Chinese Academy of Sciences, Changchun 130033, China
}
\author{Satoshi Iwamoto}
\affiliation[University of Tokyo]
{Research Center for Advanced Science and Technology, University of Tokyo, 4-6-1 Komaba, Meguro, Tokyo.}
\alsoaffiliation[University of Tokyo]
{Industrial Institute of Science, University of Tokyo, 4-6-1 Komaba, Meguro, Tokyo.}
\author{Yasutomo Ota}
\affiliation[Keio University]
{Keio University, 3-14-1 Hiyoshi, Kohoku, Yokohama, Kanagawa, Japan.}
\email{ota@appi.keio.ac.jp}

\title[Title]
  {Design of Ultrathin Faraday Rotators based on All-dielectric Magneto-optical Metasurfaces at the Telecommunication Band}

\abbreviations{MO, magneto-optical; YIG, yttrium rion garnet; BIC, bound state in the continuum; EIT, electromagnetically induced transparency; TE, transverse electric; TM, transverse magnetic.}
\keywords{Magneto-optical Effect, All-dielectric Metasurfaces, Bound State in the Continuum, Electromagnetically Introduced Transparency}

\begin{document}


\begin{center}
\textbf{Keywords:} Magneto-optical effect, all-dielectric metasurfaces, bound state in the continuum, electromagnetically induced transparency
\end{center}

\begin{abstract}
 Magneto-optical (MO) interactions offer a direct route to nonreciprocal optical devices but are intrinsically weak in the optical domain, posing a major challenge in downsizing MO functional devices. In this study, we present a design strategy for ultra-thin MO Faraday rotators based on all-dielectric metasurfaces supporting high-quality factor quasi-bound states in the continuum (QBIC) modes. Light trapping in QBIC modes induced by band folding significantly enhances MO interactions in a controllable manner, enabling a technologically relevant $45^\circ $ Faraday rotation with a MO metasurface that is only a few hundred nanometers thick. The design also incorporates electromagnetically induced transparency via spectrally overlapping resonant modes to achieve high light transmittance reaching $80\%$. This approach not only enables compact yet practical MO Faraday rotator but also holds promises for advancing free-space magnetic sensors and MO modulators. 

\end{abstract}

\section{Introduction}
Magneto-optical (MO) effects play indispensable roles in photonics as a primal approach to achieve nonreciprocity in optical devices including Faraday isolators and circulators. However, such MO devices in the optical domain are likely to be bulky due to inherently weak MO interactions in natural transparent materials\cite{MOsmallOptical,CeYIGMaterial,LowDamping} . 
To miniaturize MO devices and enable advanced applications, it is essential to enhance the effective light–matter interaction length within MO materials.

Photonic resonant structures are widely recognized for enhancing effective light-matter interaction length, thereby boosting MO effects. An earlier attempt utilized a one-dimensional photonic crystals\cite{1DMPhCreview} supporting resonant cavity modes to enhance the Faraday rotation angle ($\theta_F$)\cite{inoue1D1999,Kato1D2003,Inoue1D2016}.  However, this approach led to only modest improvement of $\theta_F$\cite{1DPhC1999,1DMPhC2defect}, and the inherently multilayered configuration tends to yield relatively thick devices\cite{1DPhCFabri,1DMPHC2018}. Plasmonic structures have also been examined for enhancing MO interactions. 
Magneto-plasmonic structures\cite{PMMSreview2016,ADaPLMSreview} based on metallic\cite{Belo-plasmon-PRL-SPP,belotelovPlasmon1-SPP,PMMS-Fedy} and hybrid metal-dielectric\cite{beloPlasmon2-PSPP,SidiskonMetal} structures have exhibited enhanced MO Kerr\cite{PMMS-Au-2022Nature,PMMS-CoAg-2024} and Faraday rotations\cite{MOFR-FoM, GoldAntenna-PSPP,SilverParticles}. However, significant optical loss caused in metals has restricted their efficiency as MO devices and thus their applications.

Among many photonic structures, all-dielectric metasurfaces are very promising due to their capability of inducing strong MO interactions within very thin and transparent structures\cite{AllDielMSMie,AlldielMSEIT,MieMsReview,ADMOMSReview2022}. Previous studies demonstrated significant enhancement of various MO responses, including circular dichroism\cite{MS-SHF,MOBIC-2,SidiskonYIG-2}, Kerr effect\cite{fabdisk,zhangGiantMagnetoopticalKerr2024,zhuDualBandKerrSensing2025} and Faraday effect \cite{ADMStoroDip-FrKr,tangEnhancingFaradayKerr2023}, through strong light confinement with high-quality ($Q$) factor toroidal dipole modes\cite{ADMStoroDip-FrKr,tangEnhancingFaradayKerr2023,ToroidalEIT,ToroidalEIT-2} and quasi-bound states in the continuum (QBIC)\cite{zhangGiantMagnetoopticalKerr2024,Zhang:24_perfect_absorb,BICMO,MOBIC-2}. Hybrid metasurfaces combining Si-based metasurfaces atop MO thin films\cite{SiDiskonYIG-BiLei2022,SiMSongarnet,SiDiskonYIGNi,gaoHybrid2023} have also been reported to enhance $\theta_F$. 
Yet, no metasurface-based free-space Faraday rotator that simultaneously achieves a high $\theta_F$ and high $T$ or a high figure of merit ($\mathrm{FoM}$ defined by $\theta_F\sqrt T$) has been reported.
To improve $T$, MO metasurfaces utilizing electromagnetically induced transparency (EIT)\cite{AlldielMSEIT,HuygensReview,Huygens,Huygensreview-3,kerker4modes,EITFR2018Chris, EnhancedFaradayRotation,gaoTilted2023} are known to be effective. However, the reported design only achieved a small $\theta_F = 7.5^\circ$ even assuming an unrealistic MO material parameter at the wavelength of the interest\cite{EITFR2018Chris}. For practical use of Faraday rotators in polarization-based isolators, $\theta_F = 45^\circ$ is necessary to maximize the forward light transmittance with maintaining the strongest isolation capability. In this sense, there needs a systematic design strategy for MO metasurfaces that realize $\theta_F = 45^\circ$ in a tunable fashion together with a high $T$. 


In this paper, we introduce a design strategy for ultra-thin MO Faraday rotators capable of simultaneously achieving $\theta_F = 45^\circ$ and high $T$. Our approach leverages high-$Q$ QBIC modes induced by band folding within an MO metasurface, enabling continuous tuning of the $Q$ factor and precise control of $\theta_F$. Remarkably, $\theta_F$ is enhanced by a factor of $1,000$ compared with that in the unprocessed host material of the same thickness. Furthermore, we reconcile $\theta_F$ tunability with high $T$ reaching $80\%$ by introducing EIT via spectral overlap of the resonance modes. This co-enhancement of $\theta_F$ and $T$ provides a practical foundation for a wide range of nanoscale MO applications.


\section{Results and Discussion}
\begin{figure}
    \centering
    \includegraphics[width=1\linewidth]{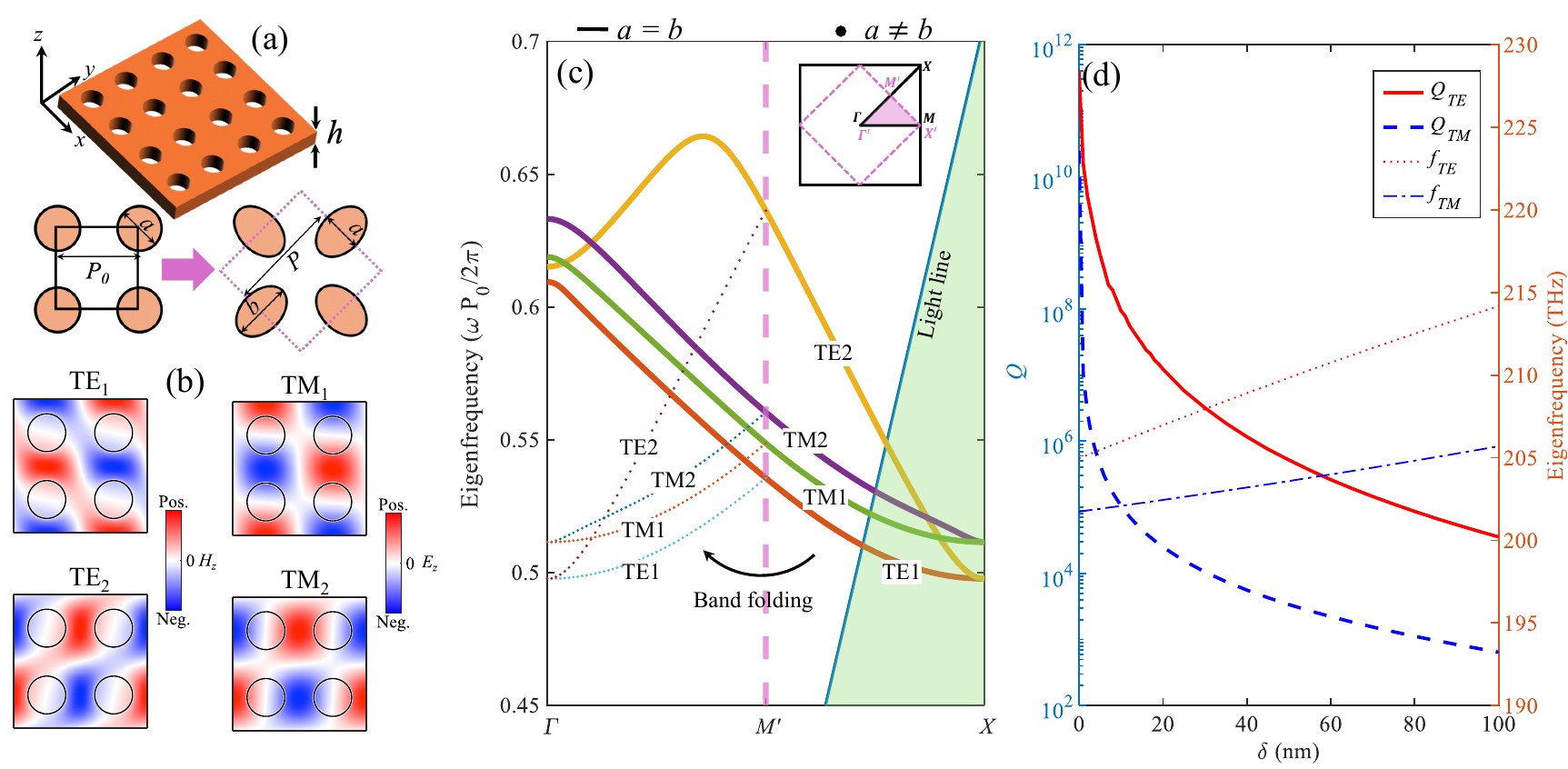}
    \caption{QBIC modes discussed in this study. (a) Conceptual illustration of the Bi:YIG based metasurfaces and the periodic perturbation approach. (b) Field distributions of investigated doubly degenerated modes at $\boldsymbol{X}$ point of the unperturbed lattice, extracted at the mid-plane of the slab, where $\mathrm{TE_1}$ and $\mathrm{TE_2}$ are represented by $H_z$ field, and $\mathrm{TM_1}$ and $\mathrm{TM_2}$ are represented by $E_z$ field. (c). Band diagram of the unperturbed lattice ($a = b = 420$ nm, $h = 400$ nm; solid lines) and the band folding of the perturbed lattice ($a = 420$ nm, $b = 421$ nm, $h = 400$ nm; dotted lines). (d) Evolution of $Q$ factors and the eigenfrequencies of TE ($f_{TE}$) and TM ($f_{TM}$) modes at the $\it\boldsymbol{\Gamma}$ point with $\delta$.}
    \label{fig:1}
\end{figure}

Our design is based on the all-dielectric MO metasurface schematically shown in Figure \ref{fig:1}a, which consists of an air-suspended square lattice of air holes in a thin membrane of bismuth-doped yttrium iron garnet (Bi:YIG). The desired Faraday rotator design will be obtained by manipulating QBIC modes found in the structure through strategically modifying the shape of the hole and thickness of the slab. The Bi:YIG layer magnetized along the $z$ direction by an external magnetic field is described by the following relative permittivity tensor:
\begin{equation}
\hat\varepsilon = 
\begin{pmatrix}
\varepsilon & -ig & 0 \\
ig & \varepsilon & 0 \\
0 & 0 & \varepsilon
\end{pmatrix}
\tag{1}
\label{equation1}
\end{equation}

where we set $\varepsilon = (2.3)^2$ and $g = 0.00235$, taken from measured values for a commercially available Bi:YIG at telecommunication wavelengths, where light absorption is negligible. The value of $g$ employed here corresponds to that of a magnetically saturated material under external magnetic field of approximately $1000$ Oe. Smaller $g$ value can be achieved by reducing the applied field using a tunable magnet; however, this approach introduces additional complexity to the device operation. Therefore, we fix $g = 0.00235$ for the following analysis. Nevertheless, as we will show shortly, our flexible design strategy enables tuning of $\theta_F$ to $45^\circ$. 


The initial structure of our design features lattice constant $P_0$, diameter of the circular airhole $a$,  slab thickness $h$, and $g = 0$. This structure supports two sets of orthogonally polarized modes, transverse electric (TE$_1$/TE$_2$) and transverse magnetic (TM$_1$/TM$_2$), which are degenerate at the $\boldsymbol{X}$ point. These modes originate from non-radiative TE-like and TM-like guided modes residing below the light line, and their field distributions at the $\boldsymbol{X}$ point are shown in Figure~\ref{fig:1}b. 
To attain radiative QBIC modes at $\it\boldsymbol{\Gamma}$ point, the structure is modified by a periodic perturbation $\delta$, which makes the airholes elliptical with a major axis length $b = a+\delta$. The perturbed lattice period becomes $P = P_0\sqrt{2}$ and is fixed at $1000$ nm. This perturbation modifies the band structures, most notably along the $\it\boldsymbol{\Gamma}$-$\boldsymbol{X}$ direction of the unperturbed structure, as shown in Figure \ref{fig:1}c. The optical bands are folded across the $\boldsymbol{M'}$ point of the perturbed structure and doubly-degenerated optical modes appear at $\boldsymbol{\Gamma}$ point. The mode degeneracies at $\it\boldsymbol{\Gamma}$ point are preserved by $C_{4V}$ symmetry. We computed $Q$ factors of the $\it\boldsymbol{\Gamma}$ point modes as a function of $\delta$ and plotted in Figure \ref{fig:1}d. We observed the divergence of $Q$ factor when approaching $\delta = 0$ and their rapid decrease with increasing $\delta$, confirming that the $\it\boldsymbol{\Gamma}$ point modes are folding-induced QBIC modes\cite{BrillouinZoneFolding}. Importantly, the changes of mode frequencies with $\delta$ are much less pronounced comparing with $Q$ factors, which allows us to control $Q$ factor nearly independently from the resonant frequencies. We note that TE-like modes consistently exhibit higher $Q$ factors than TM-like modes.

\begin{figure}
    \centering
    \includegraphics[width=0.5\linewidth]{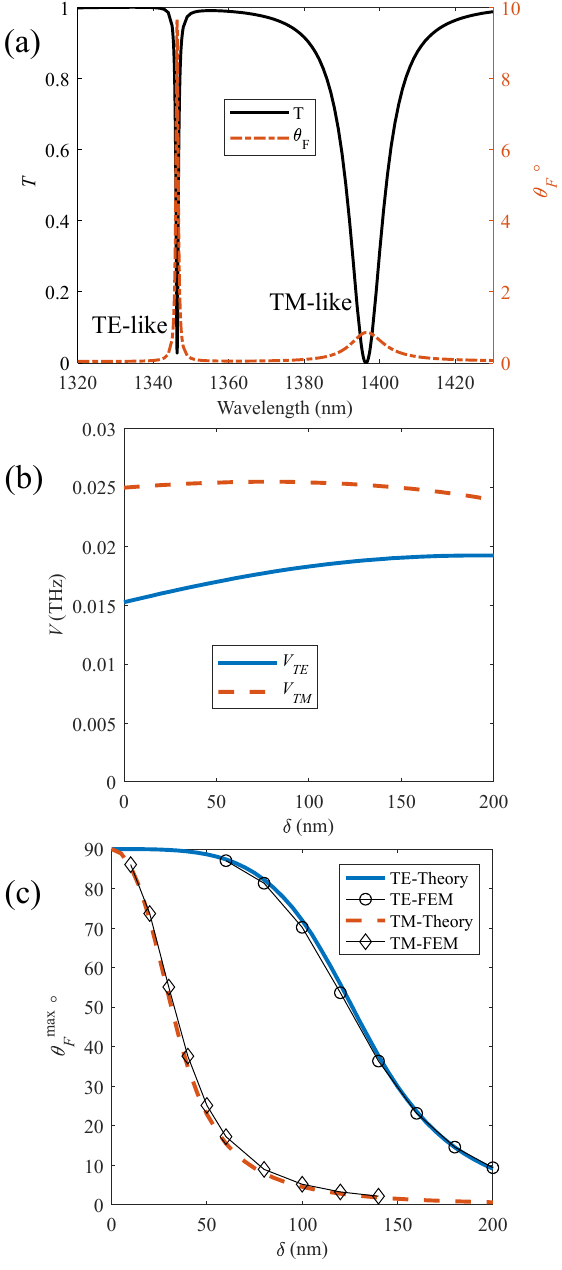}
    \caption{Light transmission and Faraday rotation of the perturbed MO metasurface. (a) Computed $T$ and $\theta_F$ in FEM simulation at $\it\boldsymbol{\Gamma}$ with $\delta = 200$ nm ($P = 1000$ nm, $a = 420$ nm, $h = 400$ nm, $g = 0.00235$); (b) Evolution of $V$ with increasing $\delta$ for TE-like and TM-like modes; (c) Theoretical and numerical results of $\theta_F^{max}$ of TE-like and TM-like modes with increasing $\delta$.}
    \label{fig:2}
\end{figure}

Subsequently, we examine light transmission and Faraday rotation at the $\it\boldsymbol{\Gamma}$ point resonant modes of the perturbed MO metasurface under an external magnetic field (i.e. $g = 0.00235$). The nonzero $g$ induces mode splitting of each set of the degenerated modes, which are now respectively turned into two orthogonal circularly polarized modes, namely the left- and right- circularly polarized (LCP and RCP) modes. 
Figure \ref{fig:2}a shows numerical simulation results for the $T$ and $\theta_F$ spectra under normal incidence of linearly polarized light with $\delta = 200$ nm and $h = 400$ nm. The transmission curves do not clearly resolve the mode splittings as they are much smaller than the spectral linewidths of the dips. We observe that the higher $Q$ factors of the TE-like modes significantly enhances $\theta_F$, reaching  $\theta_F^{TE} = 9.63^\circ$ at the resonance dip, compared to only $\theta_F^{TM} = 0.86^\circ$. 
The large $\theta_F$ observed  in the thin Bi:YIG film arises from the prolonged effective MO interaction length by the high-$Q$ factor light confinement. In the current system, the origin of Faraday rotation can be interpreted as the transmission phase difference between the two circularly polarized resonances \cite{theoryFrT}. In this context, the maximum achievable $\theta_F$ occurs at the center frequency between two MO-perturbed modes, where the phase difference between LCP and RCP modes reaches largest. The maximum rotation angle  $\theta_F^{max}$ is determined by $Q$ factor and MO-coupling strength $V$\cite{Wang:05,SHFmopertur}, as described by the following equation (see supplementary information 1):
\begin{equation}
  {\theta }_F^{max}{=}\arctan(\frac{QV}{\omega_0})
\tag{2}
\label{equation2}
\end{equation}

Where $\omega_0$ is the center frequency of the two resonances; $V$ is the spectral splitting between two perturbed modes induced by the MO effect, which is proportional to $g$ and the overlap between the MO material and modal spin density (see Supplementary Information 2). Eq. \eqref{equation2} suggests that $\theta_F^{max}$ can be flexibly controlled by adjusting $Q$ and $V$. As we discussed in Figure \ref{fig:1}d, $Q$ factors of the QBIC modes can be readily adjusted by varying the degree of the structural perturbation $\delta$. Then, we examined the influence of $\delta$ on $V$ as plotted in Figure \ref{fig:2}b. We found that $V$s for the resonant modes of interest are fairly insensitive to $\delta$ (see supplementary information Figure S2). This property is advantageous for precisely controlling $\theta_F$ to a desired value solely by adjusting $Q$ factor. Figure \ref{fig:2}c shows comparisons between simulated and theoretical $\theta_F^{max}$ as a function of $\delta$. The close agreement between theory and simulation confirms the validity of our analytical model, establishing a practical pathway for engineering $\theta_F$ in a thin Bi:YIG membrane. 

\begin{figure}
    \centering
    \includegraphics[width=0.5\linewidth]{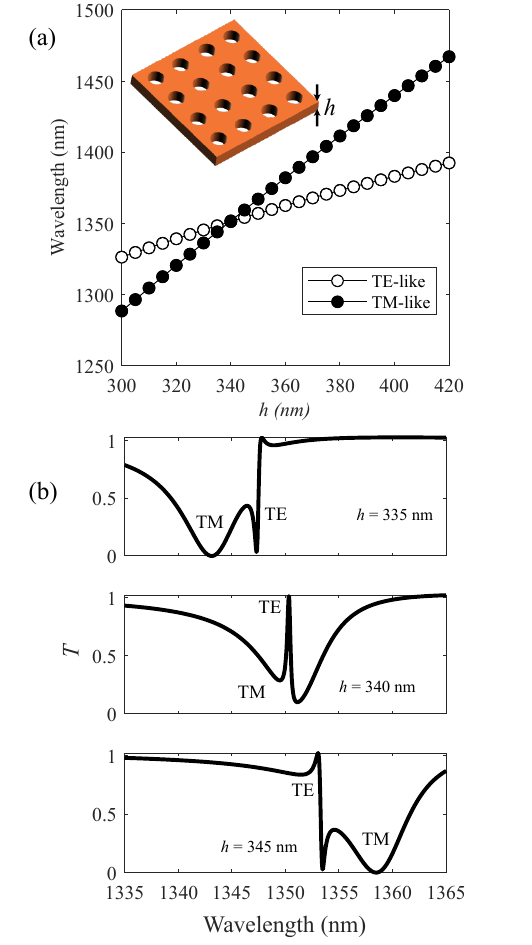}
    \caption{Realizing EIT by tunning $h$. (a) Resonant wavelengths of TE-like and TM-like modes as a function of slab thickness $h$; (b) Calculated $T$ spectra of the structure for different values of $h$. Other structural parameters are fixed at $P = 1000$ nm, $a = 400$ nm, $\delta = 200$ nm, and $g = 0$.}
    \label{fig:3}
\end{figure}

Realizing high $T$ is another important aspect in the design of practical Faraday rotators. For this purpose, we employed EIT to convert observed transmission dips of the TE-like modes into peaks by interfering with the TM-like modes. To induce EIT, the two sets of the QBIC modes are needed to be spectrally overlapped. We realize this by controlling membrane thickness $h$. Figure \ref{fig:3}a presents the resonance wavelengths of the QBIC modes as a function of $h$. To solely examine the influence of $h$, here we switch off the MO effect (i.e. $g = 0$). The resonance frequencies of the TE-like and TM-like modes differently depend on $h$ and spectrally cross near $h = 340$ nm. Figure \ref{fig:3}b shows the evolution of transmission spectra under the illumination of linearly $x$-polarized light under normal incidence. Under the non-overlapping conditions, both the TE-like and TM-like modes exhibit unwanted transmission dips. In contrast, under the resonance condition with $h = 340$ nm, the TE-like modes accompany a peak with the maximum $T$ of $99.8\%$. The realization of high $T$ peak with the TE-like modes capable of higher $\theta_F$ under nonzero $g$ is preferable for Faraday rotator design. The EIT phenomenon observed here can be understood as the forward-only positive interference between the TE-like and TM-like modes with distinct parity of the vertical radiation. We note that $Q$ and $V$ are reasonably insensitive to $h$ (see Supplementary Information Figure S3). Therefore, we can realize high $T$ simply by tuning $h$ after designing a desired $\theta_F$ of our MO metasurface, providing a firm strategy to design a practical Faraday rotator.

\begin{figure}
    \centering
    \includegraphics[width=1\linewidth]{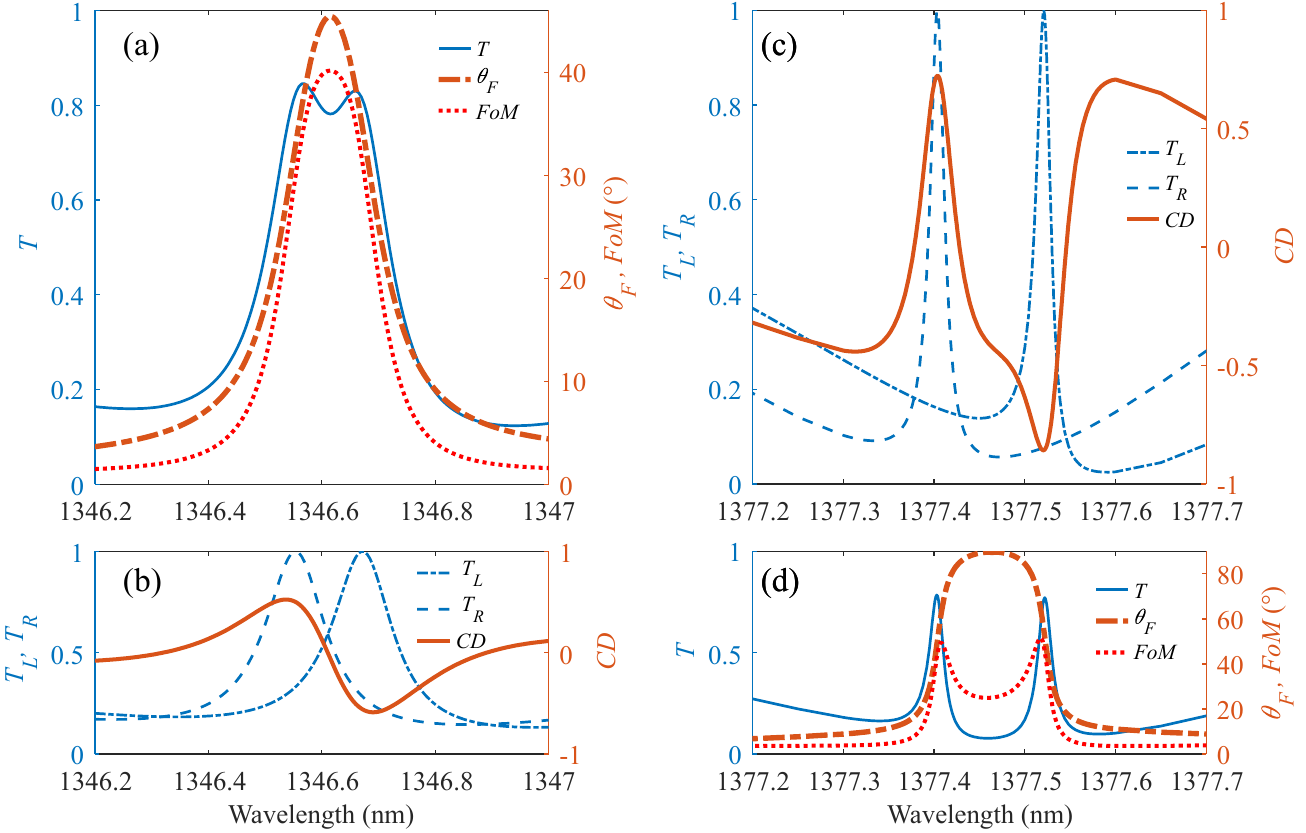}
    \caption{High $T$ and high $\theta_F$ realized in MO metasurfaces. (a) $T$, $\theta_F$, $\mathrm{FoM}$ and (b) $T_L$, $T_R$ and $CD$ for structures with perturbation $\delta = 150$ nm (Structure geometry: $P = 1000$ nm, $a = 420$ nm, $b = 557$ nm, $h = 334$ nm, $g = 0.00235$, which supported the MO-perturbed TE-like and TM-like modes with $f_{TE_L}=222.6$ THz, $f_{TE_R}=222.62$ THz, $f_{TM_L}=222.59$ THz, $f_{TM_R}=222.62$ THz) and (c), (d) for $\delta = 85$ nm (Structure geometry: $P = 1000$ nm, $a = 420$ nm, $b = 505$ nm, $h = 344.5$ nm, $g = 0.00235$ and $f_{TE_L}=217.62$ THz, $f_{TE_R}=217.64$ THz, $f_{TM_L}=217.62$ THz, $f_{TM_R}=217.65$ THz).}
    \label{fig:4}
\end{figure}

Finally, we demonstrate a high-$T$ ultra-thin Faraday rotator design with $\theta_F = 45^\circ$. As the first step of the design procedure, we estimate the required $Q$ factor for the TE-like QBIC modes ($Q_{TE}$) needed to achieve $\theta_F = 45^\circ$ at $g = 0.00235$. Accounting for the minor contribution from the TM-like modes under the EIT condition, we determine that $Q_{TE}$ need to be near $10^4$. This requirement is met by setting $\delta = 137$ nm and $h = 334$ nm, resulting in $Q_{TE} = 9.8\times 10^3$ and $Q_{TM} = 4.0\times 10^2$. By finely adjusting $h$ to $334$ nm, the TE-like and TM-like modes are spectrally overlapped, resulting in EIT at $1346.6$ nm. Figure~\ref{fig:4}a presents the $T$, $\theta_F$, and $\mathrm{FoM}$ for this optimally tuned structure under normal incidence with $x$-linearly polarized light. Sharp transmission doublet peaks originating from the TE-like modes are observed within a broad resonance dip of the TM-like modes.
Figure~\ref{fig:4}b complements the spectral responses under circularly polarized illumination, specifically the left and right circular polarization dependent transmittance, $T_L$ and $T_R$. The circularly polarized excitation reveals that the doublet arises from mode splitting due to the MO interaction. 
At the midpoint of the doublet, we obtain $\theta_F = 45^\circ$ and $T = 78\%$, corresponding to a high $FoM = 41.4$.
Moreover, the Faraday rotator exhibits low circular dichroism (CD; see Supplementary Information 5 for definition) at the peak of the Faraday rotation spectrum, where the two circularly polarized resonant modes are nearly equally excited. These results confirm the effectiveness of our design strategy under realistic material and structural parameters. 
We note that the achieved T is higher than that predicted from a simple theoretical model considering only two resonance modes\cite{theoryFrT}. We consider that this enhancement could be attributed to complex interference among multiple quasi-BIC modes involving the observed EIT phenomenon.


Figure \ref{fig:4}c further demonstrates that a design with increased $Q$ factor functions as an optical isolator exhibiting pronounced $CD$\cite{MOBIC-2}. Here, $h$ is adjusted to $345$ nm to restore EIT. These modifications yield a $Q_{TE}$ of $10^5$ while maintaining the MO coupling strength of the TE modes, leading to the accentuated splitting of the circularly polarized resonances. The light transmission through each peak is nonreciprocal and depends on the direction of light incidence. At the peak of $\lambda = 1377.4$ nm, the forward $T_L$ reaches near unity. However, the backward incidence case corresponds to $T_R$ curves and thus will show a largely reduced transmittance. The isolation performance is quantitatively represented by $CD_{max}\sim 86\%$ co-plotted in the same figure. Figure \ref{fig:4}d shows spectral responses of the structure under linearly polarized illumination. Remarkably, $\theta_F$ reaches up to $90^\circ$ between the resonance peaks. However, the value of $\mathrm{FoM}$ is limited to 26 due to the low transmittance ($T \sim 7\%$).  


\section{Conclusion}

In summary, we have developed a systematic engineering methodology for designing all-dielectric MO metasurfaces that simultaneously realize high $\theta_F$ and high $T$. Systematic tuning of structural parameters, specifically the periodic perturbation $\delta$ and slab thickness $h$, effectively control of $Q$ factors and spectral overlap of QBIC TE and TM modes. This method enabled the co-optimization of $\theta_F$ and $T$. Using these strategies, our design demonstrates $\theta_F$ up to $45^\circ$, $T$ up to $80\%$, and $\mathrm{FoM}$ exceeding $41$, corresponding to more than a $1,000$-fold enhancement in MO response compared to unstructured films of identical thickness. Access to even higher $Q$-factors enables ultra-narrow band operation, facilitating giant MO responses ($\theta_F $ up to $90^\circ$), high CD effect and pronounced optical isolation. Importantly, our design methodology is compatible with all types of transparent MO materials, establishing a broadly applicable basis for the realization of high-performance ultra-thin Faraday rotators, isolators, and magnetic sensors.

\subsection{Method}
All numerical simulations presented in this study were performed using the finite element method (FEM) implemented in COMSOL Multiphysics. The periodic metasurface was modeled as a perforated Bi:YIG slab, with periodic boundary conditions applied along both the $x$- and $y$-directions to emulate an infinite array. Perfectly matched layers (PMLs) were placed at the top and bottom boundaries to eliminate reflections. The simulation employed a two-port configuration: one port introduced a normally incident plane wave, while the other absorbed the transmitted wave. Transmittance was calculated by extracting the scattering parameters (S-parameters) from the simulation. Linear polarization, left and right circular polarizations were used for incident waves to calculate $T$, $T_L$, $T_R$. The refractive index of the surrounding medium was set to $1.0$ throughout the model.

\begin{acknowledgement}

This work was supported by JST FOREST (JPMJFR213F), JST CREST (JPMJCR19T1) and KAKENHI (25K01697, 24K17582), Mizuho Foundation for the Promotion of Science, Iketani Foundation, Nippon Sheet Glass Foundation. 

\end{acknowledgement}

\section*{Abbreviations}
MO: magneto-optical \\
YIG: yttrium iron garnet \\
BIC: bound state in the continuum \\
EIT: electromagnetically induced transparency \\
TE: transverse electric \\
TM: transverse magnetic \\

\begin{suppinfo}

A listing of the contents of each file supplied as Supporting Information
should be included. For instructions on what should be included in the
Supporting Information as well as how to prepare this material for
publications, refer to the journal's Instructions for Authors.

The following files are available free of charge.
\begin{itemize}
  \item Filename: Supplementary Information
\end{itemize}

\end{suppinfo}

\bibliography{achemso-demo}

\end{document}